\documentclass[twocolumn,aps,prl,showpacs]{revtex4}
\usepackage[T1]{fontenc}
\usepackage[latin9]{inputenc}
\usepackage{amsmath}
\usepackage{amssymb}
\usepackage{graphicx}
\usepackage{esint}

\makeatletter
\@ifundefined{textcolor}{}
{%
 \definecolor{BLACK}{gray}{0}
 \definecolor{WHITE}{gray}{1}
 \definecolor{RED}{rgb}{1,0,0}
 \definecolor{GREEN}{rgb}{0,1,0}
 \definecolor{BLUE}{rgb}{0,0,1}
 \definecolor{CYAN}{cmyk}{1,0,0,0}
 \definecolor{MAGENTA}{cmyk}{0,1,0,0}
 \definecolor{YELLOW}{cmyk}{0,0,1,0}
}

\makeatother

\begin{document}

\title{Harmonically Trapped Atoms with Spin-Orbit Coupling}

\author{Chuanzhou Zhu$^1$, Lin Dong$^1$, and Han Pu$^{1,2}$}

\affiliation{$^{1}$Department of Physics and Astronomy, and Rice Center for Quantum Materials,
Rice University, Houston, TX 77251, USA \\
$^2$Center for Cold Atom Physics, Chinese Academy of Sciences, Wuhan 430071, P. R. China}

\date{\today}
\begin{abstract}
We study harmonically trapped one-dimensional atoms subjected to an equal
combination of Rashba and Dresselhaus spin-orbit coupling induced
by Raman transition. We first examine the wave function and the degeneracy of the single-particle ground state, followed by a study of two weakly interacting bosons or fermions. For the two-particle ground state,
we focus on the effects of the interaction on the degeneracy, the spin density profiles, and the density-density correlation functions. Finally we show how these studies help us to understand the many-body properties of the system. 
\end{abstract}


\maketitle

\section{introduction}

In recent years, spin-orbit coupling and synthetic
gauge fields \cite{syn1,syn2} in ultracold atomic gases have attracted a great amount of interest. With
equal Rashba and Dresselhaus strengths, spin-orbit coupling in both
atomic Bose \cite{expb1,expb2,expb3} and Fermi \cite{expf1,expf2}
gases have been realized in experiments. Theoretically, it was predicted
that such systems would exhibit novel quantum phases for both bosonic
\cite{novel1b,novel2b,cjw1,cjw2} and fermionic \cite{novelf1,novelf2} cases,
as well as unique features such as the stripe pattern in many-body
density profile \cite{mstripe1,stripe2,novel1b}. For spin-orbit coupled systems in
uniform space, the single-particle ground state \cite{1uni} can be
easily obtained, and the ground states of two-body \cite{2uni,lin}
and many-body \cite{mstripe1} systems with zero-range contact interactions
have been analytically calculated by scattering theory and by mean field
approaches, respectively. Compared with investigating these systems in uniform space,
it may be more relevant and realistic to consider the systems in a harmonic
trap, as a trapping potential is always present in cold atom experiment. The energy spectra of trapped single-particle \cite{1trap1,1trap2,1trap3}
and two-particle \cite{Blume,Blume2} systems with spin-orbit coupling
have been numerically studied. However, a systematic investigation of the single-particle and two-particle ground states, and how they are related to the many-body physics of the trapped system is still lacking. 

In this paper we aim to present such a study. We systematically investigate the ground
states of a single particle, two bosons, and two fermions confined in a one-dimensional (1D) harmonic
trap with Raman-induced spin-orbit coupling. For the single-particle
ground state, which is presented in Sec.~\ref{sec:single-particle-ground-state}, we obtain the wave functions through imaginary time
evolution and demonstrate how the Raman coupling strength and the trap frequency affect the degeneracy. In Sec.~\ref{sec:two-boson-and-many-boson} we consider two weakly interacting bosons. The degeneracy, entanglement, density-density correlation functions, and spin density profiles of the ground
state are studied by varying the spin-dependent contact interaction, Raman coupling strength, and two-photon detuning. Our results demonstrate that the
spin-dependent interaction breaks the ground state degeneracy of this
system, and also imprints a stripe pattern in the density-density
correlation. In Sec.~\ref{sec:measure-the-gap}, we propose an experimental
scheme to measure the energy gap between the ground state and the first excited state of the system through a resonance excitation process
\cite{freqvib1,freqvib2}. In addition, the connection between the
behaviours of two-boson and many-boson ground states are discussed. To investigate the effect of quantum statistics, we then consider a system of two fermions in Sec.~\ref{sec:two-fermion-ground-state} and show how they differ from the system of two bosons. Finally, we conclude in Sec.~\ref{sec:conclusion}.

\section{single-particle ground state\label{sec:single-particle-ground-state}}

In this section, we consider a single spin-1/2 atom confined in a 1D harmonic trap with frequency $\omega$, subjected to the Raman-induced
spin-orbit coupling, with two-photon recoil momentum $q_r$, Raman coupling
strength $\Omega$, and two-photon detuning $\delta$. The Hamiltonian then takes
the form
\begin{equation}
h=\frac{\hat{p}^{2}}{2m}+\frac{1}{2}m\omega^{2}x^{2}+\frac{q_{r}\hat{p}}{m}\sigma_{z}+\frac{\Omega}{2}\sigma_{x}
+\frac{\delta}{2}\sigma_{z}\,,\label{eq:h}
\end{equation}
where $\sigma_{x}$ and $\sigma_{z}$ denote the $x$ and $z$ components of Pauli matrices, $m$
is the atomic mass, $\hat{p}=-i\hbar\partial/\partial x$ is the
momentum operator, and $x$ is the position. The two spin states are defined as 
$\sigma_{z}|\uparrow \rangle=|\uparrow \rangle$ and 
$\sigma_{z}|\downarrow \rangle=-|\downarrow \rangle$, respectively. We
mainly consider the case with $\delta=0$, and the influence of finite $\delta$
will be briefly discussed. 

\subsection{Homogeneous System}
We first briefly review the case when there is no trap \cite{1uni}, i.e.,
$\omega=0$. For this case, the system possesses translational symmetry
and thus the momentum $p$ is a good quantum number. When $\delta=0$,
the single-particle dispersion is given by
\begin{equation}
\epsilon_{p}=\frac{p^{2}}{2m}\pm\sqrt{\frac{q_{r}^{2}p^{2}}{m^{2}}+\frac{\Omega^{2}}{4}}.\label{eq:notrapE}
\end{equation}
For $\Omega<4E_{r}$ (where $E_r \equiv q_{r}^{2}/(2m)$ is the recoil energy), $\epsilon_{p}$ displays two degenerate
minima at $p=\pm k\equiv\pm q_{r}\sqrt{1-\left({\Omega}/{4E_{r}}\right)^{2}}$,
corresponding to two orthogonal degenerate ground states
\begin{align}
\left\langle x\sigma\right|g_{1}\rangle & =e^{ikx}\left[
\cos\theta_{k} \;\;\; -\sin\theta_{k} \right]^{T},\label{eq:notrapp1}\\
\left\langle x\sigma\right|g_{2}\rangle & =e^{-ikx}\left[ 
\sin\theta_{k} \;\;\; -\cos\theta_{k} \right]^{T},\label{eq:notrapp2}
\end{align}
where $\sigma=\uparrow\left(\downarrow\right)$ marks the spin up (down) state, and $\tan\theta_{k}=\frac{2}{\Omega}\left(\frac{q_{r}k}{m}+ \sqrt{\frac{q_{r}^{2}k^{2}}{m^{2}}+\frac{\Omega^{2}}{4}}\right)$.
For $\Omega>4E_{r}$, the single particle dispersion has only a single
minimum at $p=0$ and the non-degenerate ground state takes the form
\begin{equation}
\left\langle x\sigma\right|g\rangle=\left[ 
1/\sqrt{2} \;\;\; -1/\sqrt{2} \right]^{T}.\label{eq:notrapp}
\end{equation}

\subsection{Trapped System}

\begin{figure}[htp]
\includegraphics[width=8.2cm]{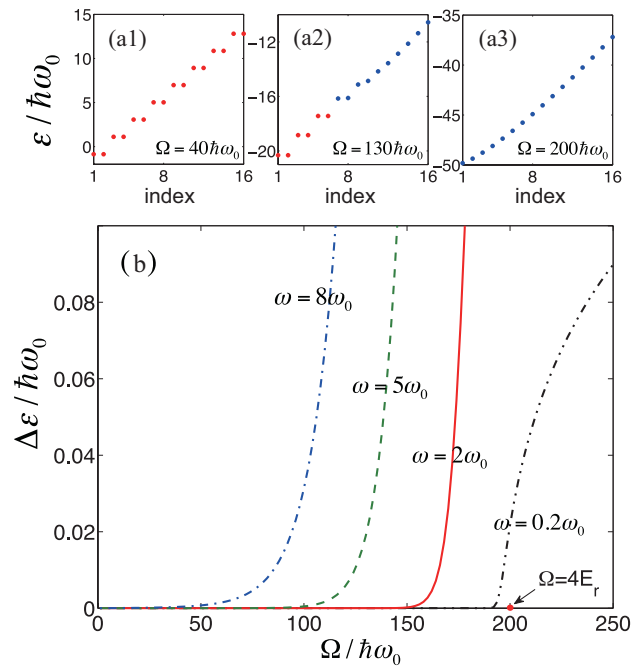}
\protect\caption{(color online) (a$_{1}$)-(a$_{3}$) The first 16 single-particle eigenenergies with trap frequency $\omega=2\omega_{0}$ for several different values of Raman coupling strength $\Omega/(\hbar\omega_{0})=40,130,200$. Red dots correspond to two-fold degenerate
eigenstates and blue dots correspond to non-degenerate eigenstates. (b) Energy gap $\Delta \epsilon$ between the two lowest energy eigenstates as a function of $\Omega$ for different values of trap frequency. We consider $\delta=0$ in this figure. Throughout this paper, we choose $\omega_{0}=2\pi\times 100\,$Hz to be the unit for frequency 
and take $m$ to be the mass of the $^{87}$Rb atom. As typical values in experiments, we choose the recoil momentum $q_{r}=10\sqrt{m\hbar\omega_{0}}$, and the trap frequency $\omega$ in the range of $0.1 \sim 10 \,\omega_{0}$.}
\label{figdE1}
\end{figure}

The presence of a harmonic trap breaks the translational symmetry
of the system, and we have to resort to numerical calculations to study the properties
of the ground state.
Using the finite difference method to discretize $\hat{p}$ and $x$,
we obtain eigenenergies through the diagonalization of the single-particle Hamiltonian (\ref{eq:h}), and the ground-state wave function by imaginary time evolution.
At $\delta=0$, Hamiltonian (\ref{eq:h}) possesses the following symmetry: If $|\psi \rangle$ is an eigenstate of $h$, $|\psi' \rangle = \sigma_x K |\psi \rangle$, where $K$ represents complex conjugate operator, is also an eigenstate with the same eigenenergy. However, unlike the time reversal symmetry of a spin-1/2 system which results in Kramer degeneracy, the current symmetry does not guarantee degenerate eigenstates. For a non-degenerate state $|\psi \rangle$, the above symmetry property necessarily requires $|\psi' \rangle = \sigma_x K |\psi \rangle$ to differ from $|\psi \rangle$ by at most an overall phase factor. In Fig.~\ref{figdE1}(a) we exhibit the low-lying energy spectrum for a trapped system with three different values of Raman coupling strength $\Omega$. Each red (blue) dot represents the energy of a two-fold degenerate (non-degenerate) eigenstate. At $\Omega=0$, we have two uncoupled spin states, and all the single-particle states must be trivially two-fold degenerate. As $\Omega$ increases, degeneracies of the high-energy states start to be lifted first. Eventually, at a critical coupling strength $\Omega_c$, the degeneracy of the ground state is also lifted and all the single particle eigenstates become non-degenerate. The energy
difference $\Delta\epsilon$ between the two lowest-lying states is shown in Fig.~\ref{figdE1}, as a function of $\Omega$
for various trap frequencies $\omega$. For a fixed value of the trap frequency $\omega$,
with increasing $\Omega$, $\Delta\epsilon$ changes from zero to finite when $\Omega$ exceeds $\Omega_c$, signaling that the ground state changes from being two-fold degenerate to non-degenerate. The critical value $\Omega_c$ at which the ground state degeneracy is lifted is a decreasing function of the trap frequency $\omega$, and in the limit $\omega \rightarrow 0$, $\Omega_c=4E_r$ and we recover the result for the homogeneous system.

The degeneracy breaking of single-particle eigenstates at large $\Omega$ can be intuitively understood as follows. The Raman coupling term, $\Omega \sigma_x/2$, in the Hamiltonian (\ref{eq:h}) behaves like an effective Zeeman field in the x-direction. At large $\Omega$, this effective Zeeman field is so strong that it polarizes the spin-1/2 particle by aligning its spin along the $x$-axis, and the particle essentially becomes a scalar particle as its spin degrees of freedom is frozen. It is a well known fact that, for a scalar particle, there is no degenerate bound state in 1D \cite{nodengeebs}.

\begin{figure}[htp]
\includegraphics[scale=0.35]{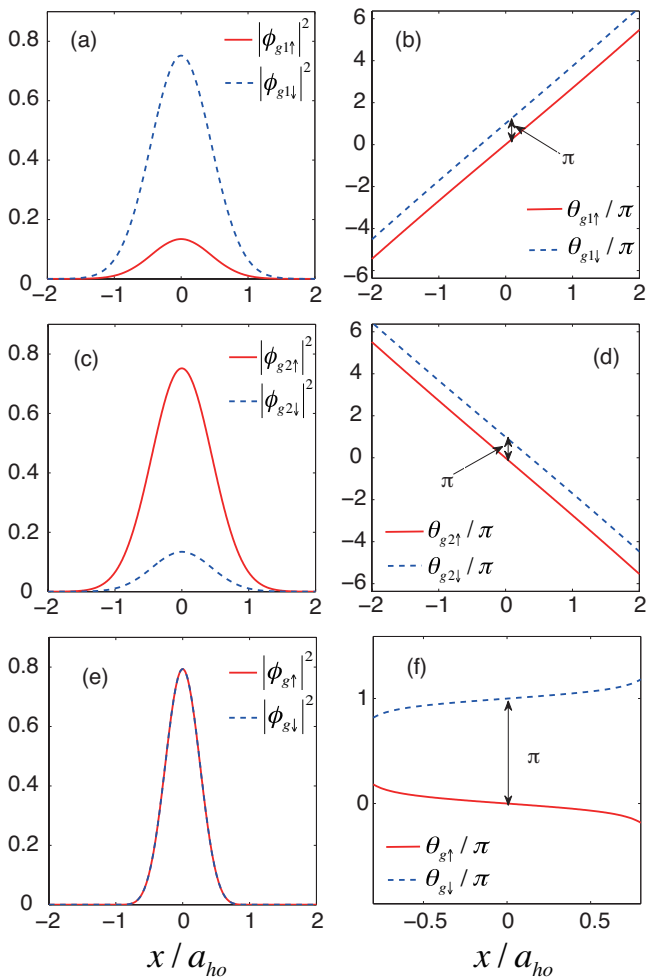}
\protect\caption{(color online) Single-particle ground state wave functions with $\omega=2\omega_{0}$,
$q_{r}=10\sqrt{m\hbar\omega_{0}}$, and $\delta=0$. Red solid line
and blue dashed line correspond to the spin-up component and the spin-down
component, respectively. For two degenerate ground states at $\Omega=130\hbar\omega_{0}$,
(a)(c) are the real space probability profiles, and (b)(d) are the
phase angles. For the non-degenerate ground state at $\Omega=200\hbar\omega_{0}$,
(e) is the real space probability profile, and (f) is the phase angle.
We define $a_{\rm{ho}}=\sqrt{\frac{\hbar}{m\omega_{0}}}$.}
\label{figp1}
\end{figure}

The two-component spinor wave
function of a single particle can be written as $\left[\begin{array}{cc}
\phi_{\uparrow}\left(x\right) & \phi_{\downarrow}\left(x\right)\end{array}\right]^{T}$, where $\phi_\sigma(x) = |\phi_\sigma(x)|\, e^{i\theta_\sigma(x)}$ is in general complex with phase angle $\theta_\sigma(x)$. In Fig.~\ref{figp1}, we exhibit the ground state wave function for $\omega=2\omega_{0}$
(which corresponds to the red solid line in Fig.~\ref{figdE1}(b)) and two different values of $\Omega$. 

When
$\Omega=130\hbar\omega_{0}$, the ground states are two-fold degenerate, and the two degenerate states are transformed to each other by the symmetry operation $\sigma_x K$. The spin density profiles for the two degenerate ground states are depicted in Fig.~\ref{figp1}(a) and (c), with the corresponding phase angles plotted in Fig.~\ref{figp1}(b) and (d), respectively. These plots suggest that we can approximately write down the ground state wave functions as 
\begin{eqnarray}
\left\langle x\sigma\right|g_{1}\rangle & = & e^{ikx}\left[\begin{array}{cc}
\phi_{1}\left(x\right) & -\phi_{2}\left(x\right)\end{array}\right]^{T} \,,\label{eq:psig1}\\
\left\langle x\sigma\right|g_{2}\rangle & = & e^{-ikx}\left[\begin{array}{cc}
\phi_{2}\left(x\right) & -\phi_{1}\left(x\right)\end{array}\right]^{T}\,,\label{eq:psig2}
\end{eqnarray}
where $\phi_{1\left(2\right)}\left(x\right)$ is real and $\left[\phi_{1\left(2\right)}\left(x\right)\right]^2$ is represented by
the red solid (blue dashed) line in Fig.~\ref{figp1}(a), and $k$
is the slope of the phase angles in Fig.~\ref{figp1}(b). The density profiles of both $\left|g_{1}\right\rangle $
and $\left|g_{2}\right\rangle $ depicted in Fig.~\ref{figp1}(a) and (c) are smooth in real space. These two states can be regarded as the analogies of the degenerate ground states for the uniform system, see Eqs.~(\ref{eq:notrapp1}) and (\ref{eq:notrapp2}). However, due to the degeneracy, any linear superposition of $|g_1 \rangle$ and $|g_2 \rangle$ represents a ground state of the system. 
Such superposition state will exhibit a
stripe pattern in its real space probability profile. 

When $\Omega=200\hbar\omega_{0}$,
the ground state is non-degenerate. This state is depicted in Fig.~\ref{figp1}(e) and (f). The symmetry property under the operation $\sigma_x K$ guarantees that, for a non-degenerate state, we must have $|\phi_\uparrow(x) |=|\phi_\downarrow (x)|$ and $\theta_\uparrow (x) + \theta_\downarrow (x) = $const. As can be seen from Fig.~\ref{figp1}(e) and (f), these conditions are indeed satisfied by the non-degenerate ground state. Furthermore, the phase angles are almost uniform. As a result, this
ground state can be approximately represented by 
\begin{equation}
 \left\langle x\sigma\right|g\rangle=\phi_0(x) \,\left[
1/\sqrt{2} \;\;\; -1/\sqrt{2} \right]^{T} \,,\label{eq:psig}
\end{equation}
where $\phi_{0}\left(x\right)$ is a real function and  $\left[\phi_{0}\left(x\right)\right]^2/2$ is plotted in Fig.~\ref{figp1}(e). This state is obviously the analogy of the non-degenerate ground state for the uniform system represented by Eq.~(\ref{eq:notrapp}).
%

Now we briefly discuss the case with finite $\delta$. In this case, the single-particle state is always non-degenerate, possessing a non-vanishing magnetization $\left\langle \sigma_{z}\right\rangle $.
In addition, the real-space wave packet of the ground state is always
smooth.

\section{two-boson ground state\label{sec:two-boson-and-many-boson}}

The two-body physics of trapped particles with spin-orbit coupling
has some non-trivial features. The research by D. Blume's group has
investigated how the real-space spin structure \cite{Blume} and the
eigenenergy spectrum \cite{Blume2} depend on $q_{r}$, $\Omega$,
and the interaction strength (In those works, the interaction has SU(2) symmetry, i.e., the interaction between different spins are characterized by the same interaction strength). Here we investigate the system from
a different perspective and focus on different parameter regimes.
We study degeneracies, density-density correlations, and density profiles
of the ground states, investigate connections between single-particle,
two-particle, and many-particle ground states, and consider the parameters
from current $^{87}$Rb experiments with a spin-dependent interaction.

In this section, we consider two weakly interacting spin-orbit coupled
bosons in a harmonic trap. To this end, we use the single-particle eigenstates discussed in the previous section to construct a
set of two-body symmetric basis vectors for expanding
the two-boson Hamiltonian. We label the
single-particle eigenstates $\left|i\right\rangle $ with corresponding eigenenergies
$\epsilon_{i}$. Then states $\left|ii\right\rangle _{b}\equiv \left|i\right\rangle _{1}\left|i\right\rangle _{2}$, and  $\left|ij\right\rangle _{b}\equiv\frac{1}{\sqrt{2}}\left(\left|i\right\rangle _{1}\left|j\right\rangle _{2}+\left|j\right\rangle _{1}\left|i\right\rangle _{2}\right)$
for $i>j$ form the symmetric two-particle basis.
Here $1,2$ are particle indices and we take a cut-off for $i,j$
in numerical calculation. Then the matrix elements of the two-particle
Hamiltonian
\begin{equation}
H=h_{1}+h_{2}+\hat{V} \,,\label{eq:Hb}
\end{equation}
with $h_{1}$ and $h_{2}$ being the single particle Hamiltonians and $\hat{V}$ the two-body contact interaction potential,
can be written as
\begin{equation}
\left\langle ij\right|\left(h_{1}+h_{2}\right)\left|kl\right\rangle _{b}=\left(\epsilon_{i}+\epsilon_{j}\right)\delta_{ik}\delta_{jl} \,,\label{eq:h1h2}
\end{equation}
and
\begin{equation}
\left\langle ij\right|\hat{V}\left|kl\right\rangle _{b}=\underset{\sigma_{1}\sigma_{2}}{\sum}g_{\sigma_{1}\sigma_{2}}\int dx \,f_{\sigma_{1}\sigma_{2}}^{ij}\left(x\right) [f_{\sigma_{1}\sigma_{2}}^{kl}\left(x\right)]^*\,,\label{eq:V}
\end{equation}
with $f_{\sigma_{1}\sigma_{2}}^{ij}\left(x\right)=_{b}\left\langle ij\right|\left.x\sigma_{1}\right\rangle _{1}\left|x\sigma_{2}\right\rangle _{2}$.
Away from the confinement induced resonance, the quasi-1D interaction
strength $g_{\sigma_{1}\sigma_{2}}$ is related to the 3D interaction strength $g_{3D\sigma_{1}\sigma_{2}}$ as  $g_{\sigma_{1}\sigma_{2}}=\frac{m\omega_{\perp}}{2\pi\hbar}g_{3D\sigma_{1}\sigma_{2}}$,
where $\omega_{\perp}$ is the strong transverse trap frequency \cite{osni}.
In our calculation, we consider a spin symmetric interaction $g_{3D\uparrow\uparrow}=g_{3D\downarrow\downarrow}=7.79\times10^{-12}$Hz cm$^{3}$,
which is from current experiments in $^{87}$Rb \cite{1uni}. 
Accordingly, we take the quasi-1D interaction strength       
$g_{\uparrow\uparrow}=g_{\downarrow\downarrow}\equiv g$ with $g=0.16\hbar\omega_{0}a_{\rm{ho}}$ where $a_{\rm{ho}}=\sqrt{\hbar/\left(m\omega_{0}\right)}$.
In the
scope of this paper, this interaction is relatively weak compared
with center-of-mass motion energy and the spin-orbit-coupling energy,
and hence the ground state of two interacting particles will not deviate
much from the ground state in the non-interacting case. To characterize the properties of the ground state, we investigate
the density-density correlation function which is defined as 
\begin{align}
C^{b}_{\sigma_{1}\sigma_{2}}\left(x_{1},x_{2}\right) & \equiv\left\langle \Psi_{g}\right|\hat{n}_{\sigma_{1}}\left(x_{1}\right)\hat{n}_{\sigma_{2}}\left(x_{2}\right)\left|\Psi_{g}\right\rangle \nonumber \\
 & =2\left|_{1}\left\langle x_{1}\sigma_{1}\right|_{2}\left\langle x_{2}\sigma_{2}\right|\left.\Psi_{g}\right\rangle \right|^{2} \,,\label{eq:Corr}
\end{align}
and the density profile which is defined as
\begin{equation}
n^{b}_{\sigma}\left(x\right)\equiv\left\langle \Psi_{g}\right|\hat{n}_{\sigma}\left(x\right)\left|\Psi_{g}\right\rangle \,, \label{eq:n}
\end{equation}
where $\hat{n}_{\sigma}\left(x\right)$ is the density operator of
spin $\sigma$ and $\left|\Psi_{g}\right\rangle$ represents
the two-boson ground state in this section.

\begin{figure}[htp]
\includegraphics[width=8.5cm]{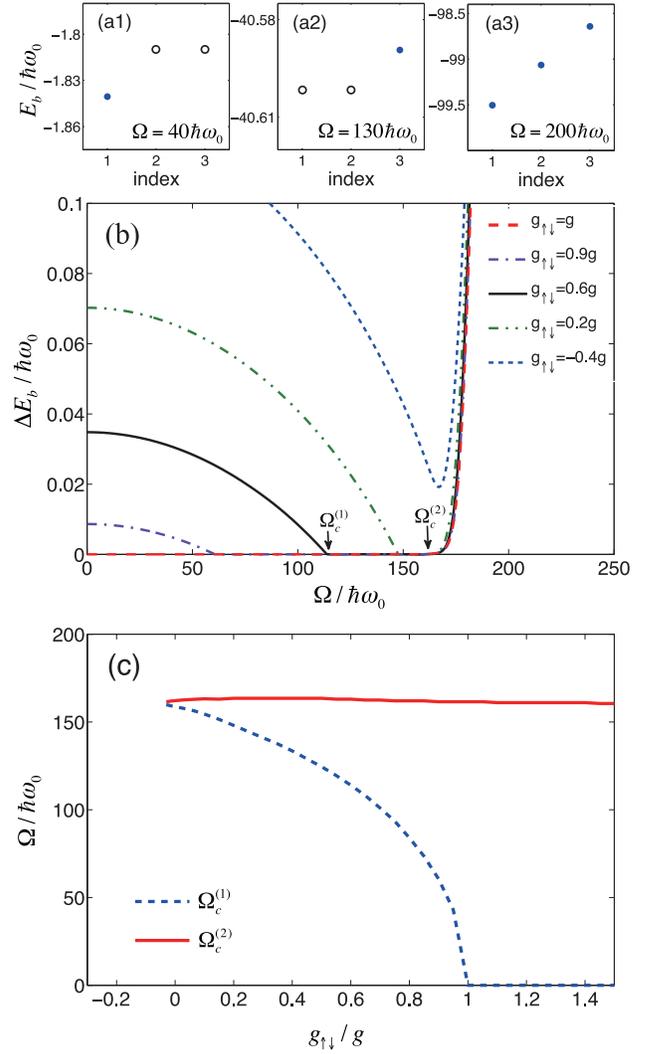}
\protect\caption{(color online) (a) Energies of the lowest three two-boson eigenstates for $g_{\uparrow\downarrow}/g=0.6$. Here solid and empty circles correspond to non-degenerate and degenerate states, respectively. (b) Energy
difference between the two lowest energy states for the case of two weakly interacting bosons, as
a function of $\Omega$, for $g_{\uparrow\downarrow}/g=1$, 0.9, 0.6, 0.2, -0.4. (c) $\Omega_{c}^{\left(1\right)}$ and $\Omega_{c}^{\left(2\right)}$
as functions of $g_{\uparrow\downarrow}$.
The other parameters are: $\delta=0$, $q_{r}=10\sqrt{m\hbar\omega_{0}}$, $\omega=2\omega_{0}$,
$g=0.16\hbar\omega_{0}a_{\rm{ho}}$. The 1D interaction
strength $g$ is calculated from 3D interaction parameter $g_{3D}=7.79\times 10^{-12}$Hz cm$^{3}$ with a transverse trapping frequency $\omega_{\bot}=100\omega_{0}$.}
\label{figdEb}
\end{figure}

\subsection{Two-body Phase Diagram at $g_{\uparrow \downarrow }= 0.6g$}

We obtain the low-lying eigenstates of the two-boson system by diagonalizing $H$ after it has been expanded onto the symmetric two-particle basis states $|ij\rangle_b$. Here we still consider the case with
$\delta=0$. From the previous study of many-boson physics, we know that the stripe phase only appears
with small $\delta$, so this regime contains the most abundant many-boson
physics. 
In the examples presented below, we choose the trap frequency to be $\omega = 2\omega_0$. We have checked that a different choice of $\omega$ does not lead to new quantum phases.
In Fig.~\ref{figdEb}(a), we plot the energies of the three lowest eigenstates for several different values of $\Omega$ with $g_{\uparrow\downarrow}=0.6g$. In Fig.~\ref{figdEb}(b), we plot $\Delta E_{b}$,
the energy difference between the two lowest energy states, as a function of $\Omega$ with several different values of $g_{\uparrow\downarrow}$.
For now let us focus on the the case with $g_{\uparrow\downarrow}=0.6g$ which is represented by the black solid line in Fig.~\ref{figdEb}(b). Depending on whether $\Delta E_b$ vanishes or not, the ground state
then exhibits the following three phases as $\Omega$ varies:

\begin{figure*}[htp]
\includegraphics[scale=0.6]{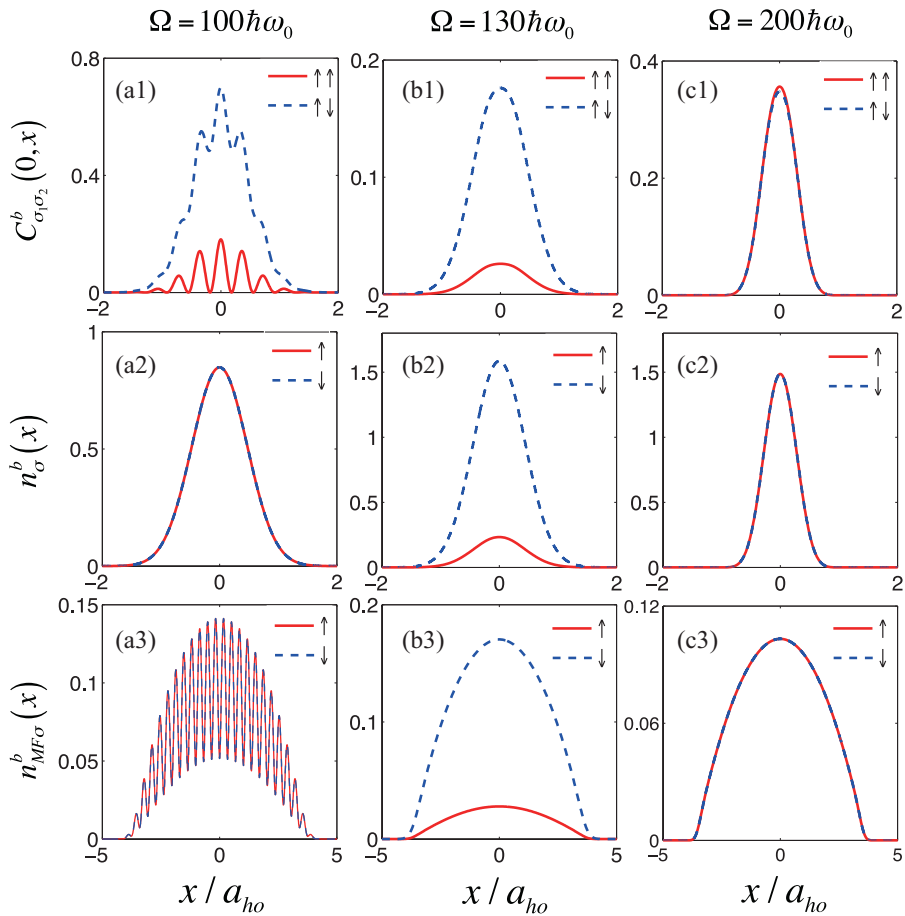}
\protect\caption{(color online) (a1)-(c1) Density-density correlation functions of two-boson ground
states.
(a2)-(c2) Spin density profiles of two-boson ground states. (a3)-(c3) Mean-field ground state
density profiles for a condensate of 1000 bosons.
The figures are plotted for the cases with $\Omega/(\hbar\omega_{0})=100,130,200$,
and $g_{\uparrow\downarrow}=0.6g$, $\delta=0$. At $\Omega=130\hbar\omega_{0}$,
the ground states are two-fold degenerate and the figures are for
one of the degenerate states. }
\label{figpb}
\end{figure*}

Phase I --- When $\Omega<\Omega_{c}^{(1)} \approx 115 \hbar \omega_0$, 
$\Delta E_{b}$ is finite, so the two-boson ground state is non-degenerate. In this regime, the single-particle ground state is two-fold degenerate and we label the two single-particle ground states as $|g_1 \rangle$ and $|g_2 \rangle$ (see discussion in Sec.~\ref{sec:single-particle-ground-state}). The ground state of the two boson system can then be approximately represented as 
\begin{equation}
\left|\text{\ensuremath{\Psi}}^{\rm{I}}_{g}\right\rangle \thickapprox\left|g_{1}g_{2}\right\rangle _{b} \equiv  \frac{1}{\sqrt{2}} \left(|g_1\rangle_1 |g_2 \rangle_2 + |g_2 \rangle_1 |g_1 \rangle_2 \right)\,.\label{tbg}
\end{equation} 
Hence the two bosons each occupies one of the single-particle ground states. Due to the bosonic statistics, the two bosons are highly entangled.
In Fig.~\ref{figpb}(a1) and (a2), we plot the 
density-density correlation $C^{b}_{\sigma_{1}\sigma_{2}}\left(0,x\right)$ and the spin density profiles $n^{b}_\sigma(x)$, respectively. The two-boson ground state features a smooth density profile identical for the two spin components. However, the entanglement manifests itself in the oscillations (or stripes) in $C^{b}_{\sigma_{1}\sigma_{2}}\left(0,x\right)$. Given the two-boson ground state in Eq.~(\ref{tbg}) and the single-particle ground states in Eqs.~(\ref{eq:psig1}) and (\ref{eq:psig2}), we can explicitly write down the density-density correlation functions and the spin density profiles as 
\begin{align}
 C^{b}_{\uparrow\uparrow}(x_{1}, x_{2})
  = & \; C^{b}_{\downarrow\downarrow}(x_{1}, x_{2}) 
 \approx 
 A_1+  B\cos\left[2k\left(x_{1}-x_{2}\right)\right] \,;  \nonumber \\
 C^{b}_{\uparrow\downarrow}(x_{1}, x_{2}) 
  = & \; C^{b}_{\downarrow\uparrow}(x_{1}, x_{2})
 \approx 
 A_2+  B\cos\left[2k\left(x_{1}-x_{2}\right)\right] \,, \nonumber \\
 n^{b}_\uparrow (x) 
  = & \; n^{b}_\downarrow(x) 
 \approx  
 \phi_1^2(x) +  \phi_2^2 (x) \,,\label{eq:Cnb1}
\end{align}
with 
\begin{align}
A_1 & \equiv\phi_{1}^{2}\left(x_{1}\right)\phi_{2}^{2}\left(x_{2}\right)+\phi_{1}^{2}\left(x_{2}\right)\phi_{2}^{2} \left(x_{1}\right);\label{eq:AB}\\
A_2 & \equiv\phi_{1}^{2}\left(x_{1}\right)\phi_{1}^{2}\left(x_{2}\right)+\phi_{2}^{2}\left(x_{1}\right)\phi_{2}^{2} \left(x_{2}\right); \nonumber \\
B & \equiv2\phi_{1}\left(x_{1}\right)\phi_{2}\left(x_{2}\right)\phi_{1}\left(x_{2}\right)\phi_{2}\left(x_{1}\right).\nonumber 
\end{align}
being smooth functions of $x_1$ and $x_2$. The stripes in the density-density correlation arise from the sinusoidal terms in Eq.~(\ref{eq:Cnb1}). We note in Fig.~\ref{figdEb}(a1) that in this phase, the first excited two-boson state is doubly degenerate. The two degenerate states roughly correspond to $|g_1 g_1 \rangle_b$ and $|g_2 g_2 \rangle_b$.

In order to connect the two-body physics to the many-body physics, we plot in Fig.~\ref{figpb}(a3) the mean-field condensate density profile. The condensate wave function is obtained by minimizing the mean-field energy functional
\begin{eqnarray}
E_{\rm MF} & =&\int dx \,\left[ N \left(
\Phi_{\uparrow}^{*} \; \Phi_{\downarrow}^{*} \right) h\left(\begin{array}{c}
\Phi_{\uparrow} \\
\Phi_{\downarrow}
\end{array}\right) \right. \label{eq:E}\\
 && \left. +\frac{N^{2}g}{2}(\left|\Phi_{\uparrow}\right|^{4}+\left|\Phi_{\downarrow}\right|^{4})+
N^{2}g_{\uparrow\downarrow}\left|\Phi_{\uparrow}\right|^{2}\left|\Phi_{\downarrow}\right|^{2} \right] \,,\nonumber 
\end{eqnarray} 
where $\Phi_{\uparrow}$ and $\Phi_{\downarrow}$ are the condensate
wave functions of the two spin components, and $N$ is the total number of atoms, which we take to be 1000 in this calculation. Fig.~\ref{figpb}(a3) shows that the condensate is in the so-called stripe phase where the density profile exhibits a stripe pattern. This stripe pattern can therefore be regarded as a manifestation of the stripes in the two-body correlation function shown in Fig.~\ref{figpb}(a1), even though the two-body density profiles are smooth. Similar connection between the mean-field many-body calculation and the quantum few-body result is also found elsewhere \cite{MFED1,MFED2}. See Appendix~\ref{app} for a more detailed discussion.


\begin{figure}[htp]
\includegraphics[width=8.5cm]{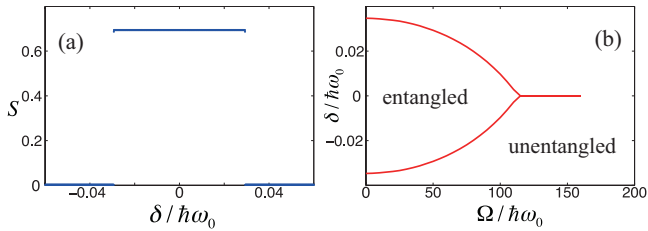}
\protect\caption{(color online) (a) Von Neumann entanglement entropy of the two-boson
ground state, as a function of $\delta$. The figure is plotted for
the case with $g_{\uparrow\downarrow}=0.6g$ and $\Omega=50\hbar\omega_{0}$.
(b) The boundary of entangled and non-entangled ground states with
$g_{\uparrow\downarrow}=0.6g$, as a function of $\delta$ and $\Omega$.
In the "entangled" region , ground states are strongly entangled and thus exhibit
stripes in density-density correlations. In the "unentangled" region, 
ground states are not entangled and show smooth density-density correlations.
Other parameters: $q_{r}=10\sqrt{m\hbar\omega_{0}}$, $\omega=2\omega_{0}$,
$g=0.16\hbar\omega_{0}$.}
\label{ent}
\end{figure}

Finally, let us consider the effect of finite two-photon detuning $\delta$. A finite $\delta$ breaks the degeneracy of the single-particle ground state. Hence one may expect that for a large $|\delta|$, the ground state of the two weakly interacting bosons corresponds to an unentangled state with both bosons occupying the non-degenerate single-particle ground state. 
For this unentangled ground state, $C^{b}_{\sigma_{1}\sigma_{2}}\left(x_{1},x_{2}\right)$ becomes smooth.
However, for a sufficiently small $|\delta|$, the two-boson ground state still roughly takes the form of Eq.~(\ref{tbg}), with $|g_1 \rangle$, $|g_2 \rangle$ representing the ground and first excited single-particle states, 
and thus $C^{b}_{\sigma_{1}\sigma_{2}}\left(x_{1},x_{2}\right)$ still exhibits stripes. 
To demonstrate the variation of the entanglement, we plot in Fig.~\ref{ent}(a) the entanglement entropy $S$ of the two-boson ground state, where $S=-{\rm Tr}[ \rho_1 \ln \rho_1]$ with $\rho_1 = {\rm Tr}_2[ |\Psi_g \rangle \langle \Psi_g|]$ being the reduced density matrix for particle 1. The state represented by (\ref{tbg}) is maximally entangled with $S=\ln 2$. Using the numerically obtained ground states with finite $|\delta|$, we find that $S$ is very close to $\ln 2$ for $|\delta| <0.03 \hbar \omega_0$, and beyond this range, $S$ quickly drops to 0, indicating an unentangled ground state. The range of $\delta$ within which the ground state is entangled is shown in Fig.~\ref{ent}(b) as a function of $\Omega$. As $\Omega$ tends to $\Omega_c^{(1)}$, this range approaches zero. 
In Fig.~\ref{ent}(b), The "entangled" ("unentangled") region manifests itself in oscillating (smooth) density-density correlations.

%
%

Phase II --- When $\Omega_{c}^{(1)}<\Omega<\Omega_{c}^{(2)}$, 
$\Delta E_{b}=0$ indicates that the ground states have two-fold degeneracy. Here $\Omega_{c}^{(2)} \approx 160 \hbar \omega_0$ is very close to the critical Raman coupling strength $\Omega_c$ at which the single-particle ground state changes from degenerate to non-degenerate.
Our result shows that the two degenerate two-boson ground states can be approximately represented as 
\begin{eqnarray} 
\label{p2}
\left|\text{\ensuremath{\Psi}}^{\rm{II}}_{g1}\right\rangle &\approx &\left|g_{1}g_{1}\right\rangle _{b}
\equiv \left|g_{1}\right\rangle_1 \left|g_{1}\right\rangle_2  \,, \label{ii} \\
\left|\text{\ensuremath{\Psi}}^{\rm{II}}_{g2}\right\rangle & \approx & \left|g_{2}g_{2}\right\rangle _{b}
\equiv \left|g_{2}\right\rangle_1 \left|g_{2}\right\rangle_2  \,. \nonumber
\end{eqnarray} 
Hence the two bosons occupy the same single-particle ground state. 
In Fig.~\ref{figpb}(b1) and (b2), we plot $C^{b}_{\sigma_{1}\sigma_{2}}\left(0,x\right)$
and $n^{b}_{\sigma}\left(x\right)$ for $\left|\text{\ensuremath{\Psi}}^{\rm{II}}_{g1}\right\rangle $,
respectively, whose explicit expressions in terms of the single-particle ground states are approximately given by:
\begin{eqnarray}
C^{b}_{\uparrow\uparrow}(x_{1}, x_{2}) & \approx &
2 \phi_1^2(x_1) \phi_1^2(x_2)\,; \label{Cnb2} \\
C^{b}_{\downarrow\downarrow}(x_{1}, x_{2}) & \approx &
2 \phi_2^2(x_1) \phi_2^2(x_2)\,; \nonumber \\
C^{b}_{\uparrow\downarrow}(x_{1}, x_{2}) & \approx & 
2 \phi_1^2(x_1) \phi_2^2(x_2)\,; \nonumber \\
C^{b}_{\downarrow\uparrow}(x_{1}, x_{2}) & \approx & 
2 \phi_1^2(x_2) \phi_2^2(x_1)\,, \nonumber \\
n^{b}_\uparrow (x) &=& 2 \phi_1^2(x) \,; \nonumber \\
n^{b}_\downarrow(x) &=& 2\phi_2^2(x) \,. \nonumber
\end{eqnarray} 
In this regime, both the density-density correlation functions and the spin density profiles are smooth functions of the position.
In addition, the total magnetization
$\mathbf{M}\equiv\int dx\left[n_{\uparrow} \left(x\right)-n_{\downarrow}\left(x\right)\right]\neq0$ in this phase. The corresponding mean-field condensate density profiles are plotted in Fig.~\ref{figpb}(b3). Here the condensate is in the so-called plane-wave phase and the density profiles for the two spin components are smooth. The condensate in this phase also exhibits finite magnetization.

Note that the two degenerate states $|\Psi^{\rm{II}}_{g1} \rangle$ and $|\Psi^{\rm{II}}_{g2} \rangle$ represented in Eq.~(\ref{ii}) are unentangled states. However, due to the degeneracy, any superposition state of $|\Psi^{\rm{II}}_{g1} \rangle$ and $|\Psi^{\rm{II}}_{g2} \rangle$ is still a ground state of the two-body system, and such a superposition state is entangled. However, this entanglement is not robust against a finite two-photon detuning $\delta$: any finite $\delta$ will force both atoms to occupy the same non-degenerate single-particle ground state, and hence destroy the entanglement and result in the smooth $C^{b}_{\sigma_{1}\sigma_{2}}\left(x_{1},x_{2}\right)$. This represents an essential difference for the ground state entanglement property between Phase I and II.

In Phase II, we note that the first excited state is non-degenerate as shown in Fig.~\ref{figdEb}(a2) and roughly corresponds to $|g_1 g_2 \rangle_b$.
Hence the ground state of Phase I corresponds to the first excited state of Phase II, and vice versa. These two phases result from the competition between the following two factors: (1) The quantum statistical property of bosons favors identical bosons to occupy the same single-particle state; and (2) the smaller inter-species interaction ($g_{\uparrow \downarrow}<g$) favors the bosons to occupy different spin states, and the smaller Raman coupling strength $\Omega$ induces more difference between the spins of $|g_1 \rangle$ and $|g_2 \rangle$. We present in Appendix \ref{App:AppendixA} a detailed and quantitative discussion on this.

Phase III --- When $\Omega>\Omega_{c}^{(2)}$, 
the gap reopens as $\Delta E_{b}$ becomes finite again. In this regime, the single-particle
ground state $|g \rangle$, whose wave function is given in Eq.~(\ref{eq:psig}), is also non-degenerate. The two-boson
ground state can then be approximately represented by 
\begin{equation}
\left|\text{\ensuremath{\Psi}}^{\rm{III}}_{g}\right\rangle \approx\left|gg\right\rangle _{b}
\equiv \left|g\right\rangle_1 \left|g\right\rangle_2
\,,
\end{equation}
which features a smooth $C_{\uparrow\uparrow}^{b}\left(0,x\right)\approx C_{\uparrow\downarrow}^{b}\left(0,x\right)$
and identical $n^{b}_\uparrow(x) = n^{b}_\downarrow (x)$, as shown in Fig.~\ref{figpb}(c1) and (c2). The explicit expressions are approximately given by:
\begin{eqnarray}
& C^{b}_{\uparrow\uparrow}(x_{1}, x_{2})
= C^{b}_{\downarrow\downarrow}(x_{1}, x_{2}) \approx 
\frac{1}{2} \phi_0^2(x_1) \phi_0^2(x_2)\,; \label{Cnb3} \\
& C^{b}_{\uparrow\downarrow}(x_{1}, x_{2})
= C^{b}_{\downarrow\uparrow}(x_{1}, x_{2}) \approx 
\frac{1}{2} \phi_0^2(x_1) \phi_0^2(x_2)\,; \nonumber \\
& n^{b}_\uparrow (x) = n^{b}_\downarrow(x) \approx \phi_0^2(x) \, \nonumber.
\end{eqnarray}
The corresponding mean-field condensate density profiles are plotted in Fig.~\ref{figpb}(c3). As in the two-body case, the condensate is smooth and features identical density profiles for the two spin components.

In this Phase, all the two-boson eigenstates are non-degenerate, as
well as all the single-particle eigenstates. The weak interaction
only causes small shifts of the eigenenergies, but does not affect
the degeneracies.

\subsection{Effects of $g_{\uparrow \downarrow}$ on Two-body Phase Diagram}

The above discussion demonstrates that, with $g_{\uparrow\downarrow}=0.6g$,
the ground state exhibits three phases separated by two critical Raman coupling strengths $\Omega_c^{(1)}$ and $\Omega_c^{(2)}$. Now let us discuss how the two-body phase diagram is changed when the inter-species interaction strength $g_{\uparrow \downarrow}$ is varied, while the intra-species interaction strength is fixed at value $g$. 

In Fig.~\ref{figdEb}(b), we also plot the energy difference between the two lowest energy states for several other values of $g_{\uparrow \downarrow}$. The dependence of $\Omega_c^{(1)}$ and $\Omega_c^{(2)}$ on $g_{\uparrow \downarrow}$ are plotted in Fig.~\ref{figdEb}(c). From these plots, we see that $\Omega_c^{(1)}$ vanishes for $g_{\uparrow \downarrow} \ge g$. In other words, when the inter-species interaction strength exceeds the intra-species interaction strength, Phase I, and hence the stripe phase in the mean-field many-body regime, no longer exists. We have checked that this property is independent of the trap frequency $\omega$.

As $g_{\uparrow \downarrow}$ decreases from $g$, $\Omega_c^{(1)}$ increases from zero and approaches $\Omega_c^{(2)}$, while $\Omega_c^{(2)}$ remains almost unchanged. Correspondingly, the parameter space where Phase II exists shrinks. At a critical value of $g_{\uparrow \downarrow}$, the two critical Raman coupling strengths merge, and for $g_{\uparrow \downarrow}$ smaller than this value, the two-body ground state is no longer degenerate for any values of $\Omega$, and $\Delta E_b$ is always positive (see the curve in Fig.~\ref{figdEb}(a) with $g_{\uparrow \downarrow}=-0.4 g$).

\subsection{Effects of $E_{r}$ and $\omega$ on Two-body Phase Diagram}
In the above discussion, we have taken the trap frequency $\omega=2\omega_{0}$
and the recoil energy $E_{r}\equiv q_{r}^{2}/2m=50\hbar\omega_{0}$,
by referring to parameters from current experiments in $^{87}{\rm Rb}$ \cite{expb1}. Since both $E_{r}$ and $\omega$ can be tuned in cold
atom experiments, here we discuss how the changes of $E_{r}$ and
$\omega$ affect the two critical Raman coupling strengths $\Omega_{c}^{\left(1\right)}$
and $\Omega_{c}^{\left(2\right)}$ for $g_{\uparrow \downarrow}=0.9g$ shown in Fig.~\ref{figdEb}(b). 

\begin{figure}[htp]
\includegraphics[scale=0.33]{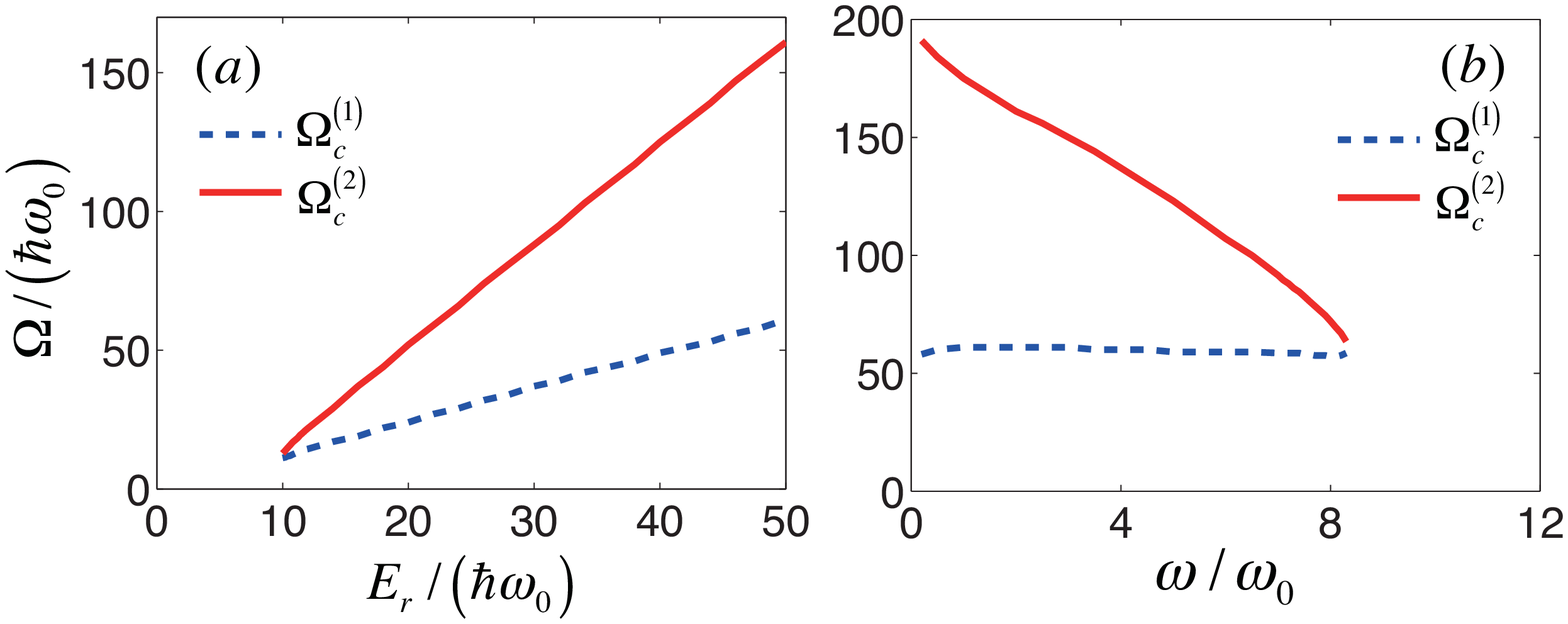}

\protect\caption{(color online) Two critical Raman coupling strengths $\Omega_{c}^{\left(1\right)}$
and $\Omega_{c}^{\left(2\right)}$, (a) as a function of recoil energy
$E_{r}$ with a fixed trap frequency $\omega=2\omega_{0}$, and (b)
as a function of trap frequency $\omega$ with a fixed recoil energy
$E_{r}=50\hbar\omega_{0}$. The other parameters are: $g_{\uparrow\downarrow}=0.9g$,
$g=0.16\hbar\omega_{0}a_{{\rm ho}}$, $\delta=0$.}

\label{fig5_2}
\end{figure}

Figure \ref{fig5_2}(a) plots the dependence of $\Omega_{c}^{\left(1\right)}$
and $\Omega_{c}^{\left(2\right)}$ on $E_{r}$ at the fixed trap frequency $\omega=2\omega_{0}$.
As $E_{r}$ decreases from $50\hbar\omega_{0}$, both $\Omega_{c}^{\left(1\right)}$
and $\Omega_{c}^{\left(2\right)}$ decrease and the regime of Phase II exists shrinks.
At a critical value of $E_{r}$ around $10\hbar\omega_{0}$, the two
critical Raman coupling strengths merge, and for $E_{r}$ smaller
than this critical value, Phase II disappears. 

Figure \ref{fig5_2}(b) depicts $\Omega_{c}^{\left(1\right)}$ and $\Omega_{c}^{\left(2\right)}$
as functions of $\omega$ at the fixed recoil energy $E_{r}=50\hbar\omega_{0}$. Here we see that
$\Omega_{c}^{\left(1\right)}$ is not very sensitive to $\omega$, while $\Omega_{c}^{\left(2\right)}$ is a decreasing
function of $\omega$. At a critical value of $\omega$, $\Omega_{c}^{\left(1\right)}$
and $\Omega_{c}^{\left(2\right)}$ merge and Phase II disappears.
That the dependence
of $\Omega_{c}^{\left(2\right)}$ on $E_r$ and $\omega$ should be expected, because the critical
$\Omega$ for the single-particle degenerate transition, which is approximately equal to $\Omega_{c}^{\left(2\right)}$, has a similar dependence on these two parameters
as shown in Fig. \ref{figdE1}(b).

\section{two-fermion ground state\label{sec:two-fermion-ground-state}}

The physics of the two-fermion ground state is quite different from
that of two bosons, because of the antisymmetric nature and the Pauli
exclusion principle for the quantum states of identical fermions. The Hamiltonian,
density-density correlation functions,
and spin density profiles of the two-fermion system are given by 
Eqs.~(\ref{eq:Hb}), (\ref{eq:Corr}), and (\ref{eq:n}), respectively, where $|\Psi_{g}\rangle$ denotes the two-fermion ground state. As we are only considering $s$-wave contact interaction, there is no intra-species interaction between two fermions. To investigate the properties of this system, we expand the Hamiltonian onto the antisymmetric two-particle basis states $|ij \rangle_f \equiv \frac{1}{\sqrt{2}}(|i \rangle_1 |j\rangle_2 - |j \rangle_1 |i \rangle_2)$, and then follow a similar procedure as above for the two-boson case. 

\begin{figure}[htp]
\includegraphics[scale=0.32]{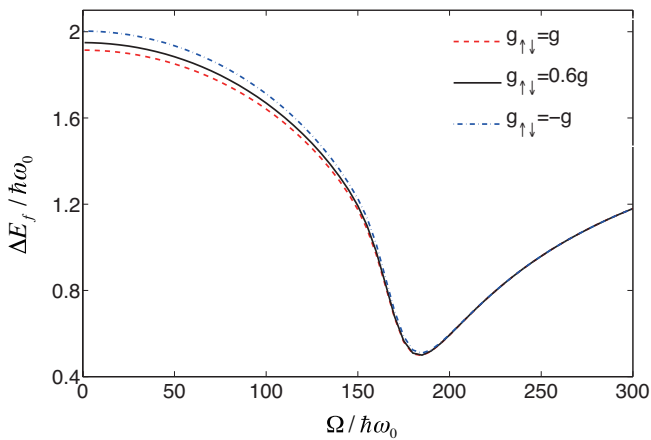}
\protect\caption{(color online) For the case of two weakly interacting fermions, energy
difference between the first excited state and the ground state, as
a function of $\Omega$ with $g_{\uparrow\downarrow}=g,0.6g,-g$.
The parameters: $\delta=0$, $q_{r}=10\sqrt{m\hbar\omega_{0}}$, $\omega=2\omega_{0}$,
$\omega_{\bot}=100\omega_{0}$, $g=0.16\hbar\omega_{0}$.}
\label{figdEf}
\end{figure}

Through exact diagonalization of the Hamiltonian in Eq.~(\ref{eq:Hb})
with $\delta=0$, we obtain $\Delta E_{f}$, the energy difference
between the two lowest-lying states, and plot it in Fig.~\ref{figdEf} as a function of $\Omega$ for several different values of $g_{\uparrow \downarrow}$. As $\Omega$ increases from zero, $\Delta E_f$ first decreases and reaches a minimum near $\Omega_c$ (the critical value of the Raman coupling strength at which the single-particle ground state degeneracy is lifted), and then starts to increase again. The essential difference with the two-boson case is that here $\Delta E_f$ is always positive and never becomes zero. Furthermore, since we are concerned
with the weak interaction regime, the interaction strength $g_{\uparrow \downarrow}$ does not have a significant effect on the system.
 
\begin{figure}[htp]
\includegraphics[scale=0.5]{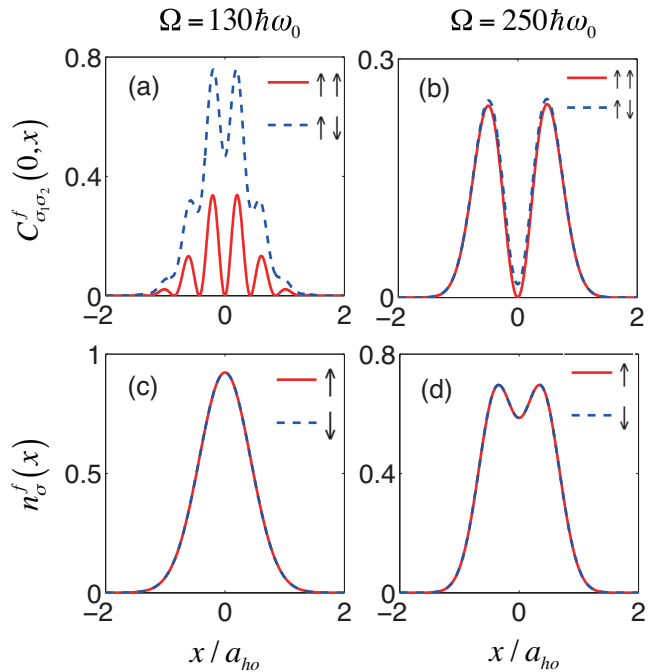}
\protect\caption{(color online) (a)(b) Density-density correlations of two-fermion
ground states, which satisfy $C_{\uparrow\uparrow}^{f}(0,x)=C_{\downarrow\downarrow}^{f}(0,x)$
and $C_{\uparrow\downarrow}^{f}(0,x)=C_{\downarrow\uparrow}^{f}(0,x)$.
(c)(d) Spin density profiles of two-fermion ground states. The figures
are plotted for the cases with $\Omega/\hbar\omega_{0}=130,250$,
$g_{\uparrow\downarrow}=0.6g$ and $\delta=0$.}
\label{figpf}
\end{figure}

In Fig.~\ref{figpf}, we display the properties of the two-fermion ground states for two Raman coupling strengths, one smaller and the other larger than $\Omega_c$:

$\Omega < \Omega_c$ --- For this case, the single-particle ground states, $|g_1 \rangle$ and $|g_2 \rangle$, are two-fold degenerate, and the two-fermion ground state can be approximately represented as 
\begin{equation}
\left|\text{\ensuremath{\Psi}}_{g}\right\rangle \thickapprox\left|g_{1}g_{2}\right\rangle _{f}=\frac{1}{\sqrt{2}} \left(|g_1\rangle_1 |g_2 \rangle_2 - |g_2 \rangle_1 |g_1 \rangle_2 \right)\,,\label{tbg1}
\end{equation} 
from which the correlation functions and density profiles can be straightforwardly calculated as 
\begin{eqnarray}
& C^{f}_{\uparrow\uparrow}(x_{1}, x_{2})
= C^{f}_{\downarrow\downarrow}(x_{1}, x_{2}) \approx 
A_1 - B\cos\left[2k\left(x_{1}-x_{2}\right)\right] \,; \nonumber \\
& C^{f}_{\uparrow\downarrow}(x_{1}, x_{2})
= C^{f}_{\downarrow\uparrow}(x_{1}, x_{2}) \approx 
A_2 - B\cos\left[2k\left(x_{1}-x_{2}\right)\right] \,;
\nonumber \\
& n^{f}_\uparrow (x) = n^{f}_\downarrow(x) \approx \phi_1^2(x) +  \phi_2^2 (x)\,, \label{eq:Cnf1}
\end{eqnarray}
with $A_1$, $A_2$, and $B$ given in Eq.~(\ref{eq:AB}). The numerical results are displayed in Fig.~\ref{figpf}(a) and (c). For this case, the density-density correlation function $C^{f}_{\sigma_{1}\sigma_{2}}(x_{1}, x_{2})$ is characterized by oscillations (or stripes) which arise from the sinusoidal terms in Eq.~(\ref{eq:Cnf1}).

$\Omega > \Omega_c$ --- For this case, the single-particle ground state $|g \rangle$ is non-degenerate. The two-fermion ground state can be approximately represented as 
\begin{equation}
\left|\text{\ensuremath{\Psi}}_{g}\right\rangle \thickapprox\left|g e\right\rangle _{f}=\frac{1}{\sqrt{2}} \left(|g\rangle_1 |e \rangle_2 - |e \rangle_1 |g \rangle_2 \right)\,,\label{tbg2}
\end{equation} 
where $|e \rangle$ denotes the non-degenerate single-particle first excited state. The density-density correlation functions and spin density profiles are displayed in Fig.~\ref{figpf}(b) and (d), respectively. 
In contrast to the single-peak structure in the previous case, here the spin density profile exhibits a double-peak structure because the real space probability profile of $|e \rangle$ features double peaks.

We remark that a system of two atoms is not unrealistic. Current technology has made it possible to trap deterministic few atoms, which allows us not only to systematically investigate the connection between few- and many-body physics, but also to study unique features of few-body systems.
In a series of experiments carried out in S. Jochim's group \cite{few}, a few-body system of fermions, with atom number precisely controlled between 1 and 10, is realized in an optical dipole trap with a fidelity of $~90\%$. If we apply the spin-orbit coupling in this kind of experiments, the ground states studied in our work should be readily obtained and their properties such as density profiles and correlations can be measured.

\section{measuring the interaction induced energy gap\label{sec:measure-the-gap}}

In Sec. \ref{sec:two-boson-and-many-boson}, we have demonstrated
that, for the two-boson case with $\delta=0$, the energy gap $\Delta E_{b}$
in Phase I with $\Omega<\Omega_{c}^{(1)}$ is induced by the spin-dependent
interaction. For fixed values of $\Omega$ and $g$ in this regime,
$g_{\uparrow\downarrow}$ and $\Delta E_{b}$ have a one-on-one mapping
relation, and hence one can obtain the value of $g_{\uparrow\downarrow}$
through measuring $\Delta E_{b}$. 

In this section, we propose an experimental scheme to measure $\Delta E_{b}$
for the two-boson case. In the $\Omega<\Omega_{c}^{(1)}$ regime,
we consider $\left|\text{\ensuremath{\Psi}}_{g}\right\rangle $ as
an initial state perturbed by a harmonic trap with a periodically modulated
trapping frequency $\omega(t)=\omega\left[1-\alpha\sin\left(\omega_{v}t\right)\right]$,
where $\omega$ is the original trap frequency, $\omega_{v}$ is the
modulation frequency, and $\alpha\ll 1$. The time evolution of the
two-boson state $\left|\text{\ensuremath{\Psi}}\left(t\right)\right\rangle $
is then determined by the Schr\"{o}dinger equation 
\begin{equation}
i\hbar\frac{\partial}{\partial t}\left|\text{\ensuremath{\Psi}}\left(t\right)\right\rangle =H_{v}\left(t\right)\left|\text{\ensuremath{\Psi}}\left(t\right)\right\rangle ,\label{eq:eqv}
\end{equation}
with the time-dependent Hamiltonian
\begin{equation}
H_{v}\left(t\right)=h_{1}^{v}\left(t\right)+h_{2}^{v}\left(t\right)+\hat{V},\label{eq:HvB}
\end{equation}
where $h_i^v(t)$ take the form of Eq.~(\ref{eq:h}) with $\omega$ replaced by $\omega(t)$.
We study the time evolution of the system by solving Eq.~(\ref{eq:eqv}) using the Crank-Nicolson method. 
\begin{figure}
\includegraphics[scale=0.3]{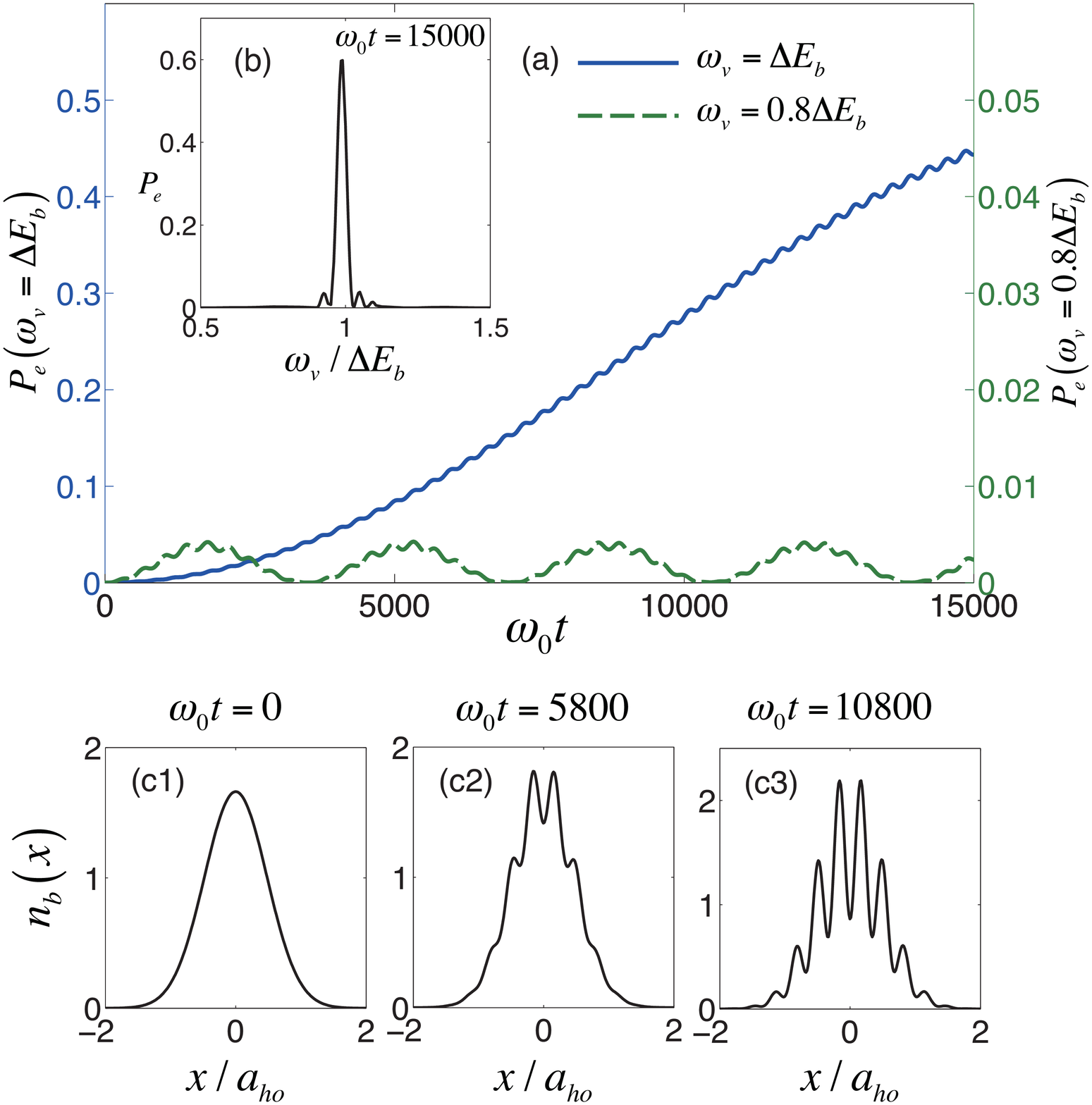}

\protect\caption{(color online) Excitation of the two-boson ground state in a harmonic
trap with periodically modulated trapping frequency $\omega\left[1-\alpha\sin\left(\omega_{v}t\right)\right]$.
We fix $g_{\uparrow\downarrow}=0.6g$, $\Omega=80\hbar\omega_{0}$,
$\omega=2\omega_{0}$, $\alpha=0.05$, and choose the two-boson ground
state as the initial state. (a) Probability $P_{e}\left(t\right)$
on the first excited state for the on-(off-)resonance case with $\omega_{v}=\Delta E_{b}\left(0.8\Delta E_{b}\right)$,
as a function of time $t$. (b) $P_{e}$ as a function of $\omega_{v}$
when $\omega_{0}t=15000$. (c1)-(c3) Time evolution of the total density
profile $n_{b}\left(x\right)$ for the on-resonance case with $\omega_{v}=\Delta E_{b}$.}

\label{fig8}
\end{figure}

The measurement of $\Delta E_{b}$ can be conducted by making use
of the resonant excitation of the system. We investigate how periodic
perturbations with various $\omega_{v}$ influence the probability
$P_{e}\left(t\right)$ for the ground state to be excited to the first
excited state. 
In the following discussion, we consider $g_{\uparrow\downarrow}=0.6g$
and $\Omega=80\hbar\omega_{0}$, and hence
define $\Delta E_{b}\equiv\Delta E_{b}\left(\Omega=80\hbar\omega_{0}\right)$. For an on-resonance modulation with $\omega_v=\Delta E_b$, we see in Fig.~\ref{fig8}(a) a significant growth of $P_e$, whereas for an off-resonance modulation with $\omega_v=0.8\Delta E_b$, $P_e$ never exceeds $0.5\%$. 
We plot in Fig.~\ref{fig8}(b) the excitation probability $P_{e}$ as
a function of the modulation frequency $\omega_{v}$
at $\omega_{0}t=15000$, where a typical resonance peak is clearly seen. 

In order to visualize the above resonant excitation process, we examine
the time evolution of the total density profile which is defined as
$n_{b}\left(x\right)\equiv n_{\uparrow}^{b}\left(x\right)+n_{\downarrow}^{b}\left(x\right)$.
For the on-resonance case with $\omega_{v}=\Delta E_{b}$, $n_{b}\left(x\right)$
develops a stripe pattern as a function of $t$, as shown in Figs.~\ref{fig8}(c1)-(c3). The presence of this stripe pattern is because
$\left|\text{\ensuremath{\Psi}}\left(t\right)\right\rangle _{b}$
becomes a superposition of the ground state $\left|\text{\ensuremath{\Psi}}_{g}\right\rangle $
and the first excited state 
during the time evolution. However, for off-resonant modulation, the
system is almost unaffected by the periodic perturbation, and the
stripe pattern is not present in $n_{b}\left(x\right)$.

\section{conclusion\label{sec:conclusion}}

In this paper we have systematically investigated the single-particle and two-body ground states of
Raman-induced spin-orbit coupled ultracold atoms in a 1D harmonic trap. In the absence of the Raman coupling, all single-particle eigenstates are two-fold degenerate. As the Raman coupling strength increases, the degeneracy of higher energy eigenstates start to be lifted first, and eventually at a critical coupling strength, the ground state (and hence all energy eigenstates) becomes non-degenerate. The single-particle spectrum and wave functions help us to understand the two-body properties of the system for both bosons and fermions. For the two-boson case, we point out three phases distinguished by the behaviors of the degeneracy, density-density correlation functions, and spin density profiles. Then we identify a regime where the two atoms in the ground state are entangled and  characterized by stripes in density-density correlations. This regime corresponds to the regime of the exotic stripe phase in the mean-field many-body limit. Our work therefore establishes a connection among one-, few- and many-body physics of trapped atomic systems with spin-orbit coupling.

\begin{acknowledgments}
This work is supported by NSF and the Welch Foundation (Grant No.
C-1669).
\end{acknowledgments}

\appendix

\section{Two-body vs. Mean-Field Many-Body results for bosons} \label{app}
One of the goals of our work is to bridge the two-body and the many-body physics. Most of the many-body properties of weakly interacting condensate can be well understood under the mean-field framework. In the mean-field approach, the underlying assumption is that all the atoms occupy the same single-particle state and quantum entanglement between atoms is neglected. The two-body ground states in Phase II and III studied in Sec.~\ref{sec:two-boson-and-many-boson} are consistent with this assumption, and the connection between the two-body and many-body physics can be easily seen from the right two columns of Fig.~\ref{figpb}. By contrast, Phase I requires some special attention.

In Phase I of the two-boson system, the single-particle ground states are two-fold degenerate and are denoted as $|g_1 \rangle$ and $|g_2 \rangle$, and the two-boson ground state is given in Eq.~(\ref{tbg}), which we rewrite here:
\begin{equation}
\left|\text{\ensuremath{\Psi}}^{\rm{I}}_{g}\right\rangle \thickapprox\left|g_{1}g_{2}\right\rangle _{b} \equiv  \frac{1}{\sqrt{2}} \left(|g_1\rangle_1 |g_2 \rangle_2 + |g_2 \rangle_1 |g_1 \rangle_2 \right)\,.
\end{equation} 
State $\left|\text{\ensuremath{\Psi}}^{\rm{I}}_{g}\right\rangle$ is a maximally entangled state, hence is expected to be very different from the mean-field ground state. In fact, in this regime, in the language of second quantization the mean-field many-body ground state can be roughly represented as 
\begin{equation} 
|\Psi _{\rm MF}(\theta) \rangle = \frac{1}{\sqrt{2^N N!}} \left( e^{-i\theta/2} a_1^\dag + e^{i\theta/2} a_2^\dag \right)^N |0 \rangle \,,\label{MF}
\end{equation} 
where $N \gg 1$ is the number of particles, $|0 \rangle$ is the vacuum state with no atoms, and $a_i^\dag$ is the creation operator that create an atom in single-particle state $|g_i \rangle$. This mean-field state corresponds to the situation where all atoms occupy the same single-particle state $(e^{-i\theta/2} |g_1 \rangle + e^{i\theta/2} |g_2 \rangle) /\sqrt{2}$ which is a coherent superposition of the two single-particle ground state, with $\theta$ being an arbitrary phase. This state possesses no quantum entanglement between different particles, but do give rise to the density stripe as shown in Fig.~\ref{figpb}(a3). 

We can draw an analogy from a different system: a system of many scalar bosons in a double-well potential, in which the operators $a_i^\dag$ correspond to creation operator that creates a particle in the $i$th well ($i=1$, 2). This system is analyzed in detail in Ref.~\cite{MFED1}. The mean-field analysis yields a state similar to Eq.~(\ref{MF}), but the quantum calculation produces a different result to the mean field.

The mean-field state $|\Psi_{\rm MF}(\theta) \rangle$ in Eq.~(\ref{MF}) has, on average, $N/2$ atoms in both $|g_1 \rangle$ and $|g_2 \rangle$. However, the occupation number for each of these states possess large fluctuations. It is probably beyond anyone's capability to write down the full quantum many-body wavefunction for this system. However, we can still make some qualitative remarks with insights drawn from Ref.~\cite{MFED1}. In quantum treatment, large number fluctuations, as included in the mean-field state $|\Psi_{\rm MF}(\theta) \rangle$, are in general not favored by interaction, which tends to drive the state into the Fock state:
\begin{equation}
|\Psi_{\rm F} \rangle = \frac{1}{(N/2)!} (a_1^\dag)^{N/2} (a_2^\dag)^{N/2}\,  |0 \rangle \,,
\end{equation}
which is just the $N$-body analog of the two-body ground state $|\Psi_g^{\rm I} \rangle$. Furthermore, the Fock state $|\Psi_{\rm F} \rangle$ may be roughly regarded as a superposition of mean-field states averaged over the phase $\theta$, i.e., 
\[ |\Psi_{\rm F} \rangle  \approx \emph{C} \int_{0}^{2 \pi} d\theta \,|\Psi _{\rm MF}(\theta) \rangle \, \] where $\emph{C}$ is a normalization constant. Conversely, the mean-field state may be regarded as a broken-symmetry state with a random but fixed $\theta$.

In summary, we can establish the following connection between the two-body results and the mean-field many-body results for Phase I. The two-body ground state does not exhibit strips in the density profiles, as can be seen in Fig.~\ref{figpb}(a2), but does contain quantum entanglement between the particles and exhibit oscillations in the correlation function, as shown in Fig.~\ref{figpb}(a1). As the number of atoms increases and the mean-field limit is approached, quantum entanglement becomes more and more fragile, and is completely neglected in the mean-field treatment. The mean-field assumption is also consistent with the picture of spontaneous symmetry breaking, where a random but specific $\theta$ is selected and all the atoms are considered to be condensed into the linear superposition state of $|g_1 \rangle$ and $|g_2 \rangle$ with a phase difference $\theta$. Such a state leads to the stripe pattern in the density profile, as shown in Fig.~\ref{figpb}(a3). We can therefore state the following: For the system under current study, through spontaneous symmetry breaking, the oscillations in two-body correlation function become manifest in the stripes of mean-field density profiles.

We remark that this connection between few-body correlation function and mean-field density profile is not unique to our system. For example, Kanamoto {\em et al.} studied a system of attractive scalar bosons confined along a ring \cite{MFED2}. When the attractive interaction strength exceeds a critical value, the mean-field calculation shows that the density profile of the BEC becomes inhomogeneous and take the form of a bright soliton. However, the quantum calculation for a few-body system always gives a homogeneous density profile, but the second-order correlation function exhibit inhomogeneity.

\section{Two-Boson Ground State in Phase I and Phase II} \label{App:AppendixA}

In both Phase I and II of the two-boson system, the single-particle ground states are two-fold degenerate and are denoted as $|g_1 \rangle$ and $|g_2 \rangle$. In the case of two interacting bosons, the following two situations represent potential candidates for the ground state:
\begin{enumerate}
\item Both atoms occupy the same single-particle ground state. Hence the two-boson state is given by $|g_1 g_1 \rangle_b$ or $|g_2 g_2 \rangle_b$, which are the states represented by Eqs.~(\ref{p2}). From symmetry, we know that these two states are always energetically degenerate, hence we only consider $|g_1 g_1 \rangle_b$ in the following.
\item The two atoms occupy different single-particle ground state. Hence the two-boson state is given by $|g_1 g_2 \rangle_b$, which is the state represented by Eq.~(\ref{tbg}).
\end{enumerate}
The question of concern is which state, $|g_1 g_1 \rangle_b$ or $|g_1 g_2 \rangle_b$, possesses the lower energy. It is obvious that we only need to compare the interaction energy associated with these two states, which we denote as $E^{\rm I, II}_{\rm int}$ with
\begin{eqnarray*}
 E^{\rm I}_{\rm int} &\equiv & _b\langle g_1 g_2 |\hat{V} |g_1 g_2 \rangle_b \,,\\
 E^{\rm II}_{\rm int} &\equiv & _b\langle g_1 g_1 |\hat{V} |g_1 g_1 \rangle_b \,.
\end{eqnarray*}

With Eqs.~(\ref{eq:psig1}),~(\ref{eq:psig2}), and~(\ref{eq:V}), it is straightforward to show that
\begin{eqnarray*}
E^{\rm I}_{\rm int}  & =& \left(4g+2g_{\uparrow\downarrow}\right)D+g_{\uparrow\downarrow}F\,,\label{eq:A1}\\
 E^{\rm II}_{\rm int}   & = & 2g_{\uparrow\downarrow}D+gF \,,\nonumber 
\end{eqnarray*}
where
\begin{align}
D & =\int dx \,\phi_{1}^{2}\left(x\right)\phi_{2}^{2}\left(x\right)\,,\\
F & =\int dx \, \left[\phi_{1}^{4}\left(x\right)+\phi_{2}^{4}\left(x\right)\right]\,,\nonumber 
\end{align}
from which we have \begin{equation}
\label{com}
E^{\rm I}_{\rm int} - E^{\rm II}_{\rm int} = gF \left[\frac{g_{\uparrow \downarrow}}{g}-f(\Omega) \right] \,, 
\end{equation}
where $f(\Omega) = 1-4D/F$ and, according to our numerics, is a decreasing function of $\Omega$ and satisfies the condition $0 < f(\Omega) < 1$. 

If the interaction is spin-independent, i.e., $g_{\uparrow \downarrow} = g$, Eq.~(\ref{com}) shows that $E^{\rm I}_{\rm int} > E^{\rm II}_{\rm int}$, the two-boson system is in Phase II and the two atoms occupy the same single-particle state. This is the manifestation of the bosonic statistics, which favors repulsive bosons to occupy the same state. 

For spin-dependent interaction, the situation is more complicated. If $g_{\uparrow \downarrow} >g$, $E^{\rm I}_{\rm int} > E^{\rm II}_{\rm int}$ still holds, and the interaction also favors two atoms occupying the same state. Hence the interaction effect and the statistical property strengthens each other, and the system remains in Phase II. If $g_{\uparrow \downarrow} < g$, the interaction favors two atoms occupying different states, and hence competes with the statistical effect. For a given $\Omega$, when $g_{\uparrow \downarrow} < gf(\Omega)$, the interaction wins the competition and the system enters Phase I. The above analysis also agrees with Fig. \ref{figdEb}(b) by showing that for a given $g_{\uparrow \downarrow} < g$, the system enters Phase I when $\Omega < f^{-1}(g_{\uparrow \downarrow}/g)$.

One may still ask the question: In Phase II, can the two atoms occupy the same single-particle state which is a linear superposition of $|g_1 \rangle$ and $|g_2 \rangle$? For example, how about the state $|\Psi_{\rm MF}(\theta) \rangle$ as in Eq.~(\ref{MF}) with $N=2$,
which represents the state where both atoms occupy the single-particle state $(e^{-i\theta/2} |g_1 \rangle+e^{i\theta/2}|g_2 \rangle)/\sqrt{2}$. The answer is that such a state is not favored in Phase II. This can also be understood from the interaction energy. State $|\Psi_{\rm MF}(\theta) \rangle$ can be regarded as a linear superposition of $|g_1 g_1 \rangle_b$, $|g_2 g_2 \rangle_b$, and $|g_1 g_2 \rangle_b$. The first two are the degenerate two-boson ground state in Phase II, while the last one is the corresponding first excited state. Hence $|\Psi_{\rm MF}(\theta) \rangle$ cannot represent the ground state.

\end{document}